\documentclass[10pt,dvipdfm]{article}
\usepackage{amsmath,epsfig,epsf,psfrag, amssymb,amsfonts,amsbsy, bbding, pifont, fancyhdr,bm}

\usepackage{booktabs,dcolumn}

\newcommand{\Lower}[1]{\smash{\lower 2ex \hbox{#1}}}
\usepackage[mathscr]{eucal}
\usepackage{color}
\usepackage[bf]{titlesec}
\usepackage{hyperref}
\usepackage{rotating}
\usepackage{cite}
\usepackage{natbib}

\date{}

\newcommand{\BE}{\begin{equation}}
\newcommand{\EE}{\end{equation}}

\newcommand{\MM}{{\mathcal M}}

\newcommand{\bmzero}{\mbox{\bf 0}}

\newcommand{\bmb}{\mbox{\boldmath $b$}}

\newcommand{\bme}{\mbox{\boldmath $e$}}

\newcommand{\bmy}{\mbox{\boldmath $y$}}

\newcommand{\bmB}{\mbox{\bf B}}

\newcommand{\bmE}{\mbox{\boldmath $E$}}

\newcommand{\bmI}{\mbox{\bf I}}

\newcommand{\bmR}{\mbox{\boldmath $R$}}
\newcommand{\bmS}{\mbox{\bf S}}

\newcommand{\bmU}{\mbox{\boldmath $U$}}
\newcommand{\bmV}{\mbox{\boldmath $V$}}

\newcommand{\bmY}{\mbox{\boldmath $Y$}}
\newcommand{\bmZ}{\mbox{\boldmath $Z$}}

\newcommand{\bmmu}{\mbox{\boldmath $\mu$}}

\newcommand{\bmtheta}{\mbox{\boldmath $\theta$}}
\newcommand{\bmTheta}{\mbox{\boldmath $\Theta$}}

\newcommand{\bmGamma}{\mbox{\boldmath $\Gamma$}}

\newcommand{\bmOmega}{\mbox{\boldmath $\Omega$}}
\newcommand{\bmtau}{\mbox{\boldmath $\tau$}}

\newcommand{\bmPsi}{\mbox{\boldmath $\Psi$}}

\newcommand{\bmvarepsilon}{\mbox{\boldmath $\varepsilon$}}
\newcommand{\bmSigma}{\mbox{\boldmath $\Sigma$}}

\def\undertilde#1{\mathord{\vtop{\ialign{##\crcr
   $\hfil\displaystyle{#1}\hfil$\crcr\noalign{\kern1.5pt\nointerlineskip}
   $\hfil\tilde{}\hfil$\crcr\noalign{\kern1.5pt}}}
}}

\newcommand{\Yaug}{\mbox{$\bmY_{\hbox{aug}}^{[1]}$}}
\newcommand{\Yaugtwo}{\mbox{$\bmY_{\hbox{aug}}^{[2]}$}} 
\newcommand{\Yaugthree}{\mbox{$\bmY_{\hbox{aug}}^{[3]}$}} 
\def\thesection{\arabic{section}}
\def\thesubsection{\arabic{section}.\arabic{subsection}}
\titleformat{\section}[hang]{\bf\large}{\thesection}{0.25cm}{}
\titleformat{\subsection}[hang]{\normalsize}{\thesubsection}{0.25cm}{}

\newcolumntype{.}[1]{D{.}{.}{#1}}
\begin{document}
\title{Capturing Patterns via Parsimonious $t$ Mixture Models}
\date{}
\author{Tsung-I~Lin$^{12}$\thanks{Corresponding author. E-mail:~{tilin@amath.nchu.edu.tw}; tel.:+886 4 22850420; fax: +886 4 22873028.
},~~Paul~D.~McNicholas$^{3}$~~and~~Hsiu~J.~Ho$^{1}$\\
\footnotesize\it $^{1}$Institute of Statistics, National Chung Hsing University, Taichung 402, Taiwan,\\
\footnotesize\it $^{2}$Department of Public Health, China Medical University, Taichung 404, Taiwan\\
\footnotesize\it $^{3}$University Research Chair in Computational Statistics at the University of Guelph, Ontario, Canada\\
}

\maketitle

\begin{abstract}
This paper exploits a simplified version of the mixture of multivariate $t$-factor analyzers (MtFA)
for robust mixture modelling and clustering of high-dimensional data that frequently contain a number of outliers.
Two classes of eight parsimonious $t$ mixture models are introduced and computation of maximum likelihood
estimates of parameters is achieved using the alternating expectation conditional maximization
(AECM) algorithm. The usefulness of the  methodology
is illustrated through applications of image compression and compact facial representation.
\end{abstract}
{\em Keywords:}~~Factor analysis; Facial representation; Image compression; PGMM; PTMM


\section{Introduction}
The mixture of factor analyzers (MFA) model, first introduced by  \cite{ghahramani97}
and further generalized by 
\cite{mclachlan00a}, 
has provided a flexible dimensionality reduction approach to the
statistical modelling of high-dimensional data arising from  a
wide variety of random phenomena. By combining the
factor analysis model with
finite mixture models, the MFA model allows simultaneous partitioning of the
population into several subclasses while performing a local
dimensionality reduction for each mixture component. In recent
years, the study of MFA  has received considerable interest; see
\cite{mclachlan00b} and \cite{fokoue03}
for some excellent reviews. \cite{mclachlan02} and \cite{mclachlan03} exploited the MFA approach to
handle high-dimensional data such as clustering of microarray
expression profiles. \cite{mcnicholas08} generalized the
MFA model by introducing a family of parsimonious Gaussian mixture models (PGMMs).

In the MFA framework, component errors and factors are
routinely assumed to have a Gaussian distribution due to their
mathematical tractability and computational convenience. In
practice, noise components or badly discrepant outliers often
exist; the multivariate $t$ (MVT) distribution contains an
additional tuning parameter, the degrees of freedom (df), which can be useful
for outlier accommodation. Specifically, the density of a $g$-component multivariate $t$ mixture model is of the form
\begin{equation}
f(\bmy)=\sum_{i=1}^gw_it_p(\bmy|\bmmu_i,\bmSigma_i,\nu_i),\label{tmix}
\end{equation} where the $w_i$ are mixing proportions and $t_p(\bmy|\bmmu_i,\bmSigma_i,\nu_i)$ is the density of
$p$-variate $t$ distribution with df $\nu_i$,
mean vector $\bmmu_i$, and scaling covariance matrix $\bmSigma_i$. Note that the
general $t$ mixture (TMIX) model (\ref{tmix}) has a total of
$(g-1)+gp+gp(p+1)/2+g$ free parameters, of which $gp(p+1)/2+g$ parameters correspond to the
component matrices $\bmSigma_i$ and df $\nu_i$, respectively. In a more parsimonious version of (\ref{tmix}), wherein the $\bmSigma_i$ and $\nu_i$ are restricted to be identical across groups, the number of free
parameters reduces to $(g-1)+gp+g(p+1) + p(p + 1)/2$.

The TMIX model was first considered by \cite{mclachlan98} and \cite{peel00},
who presented expectation-maximization (EM)
algorithms for parameter estimation and show the
robustness of the model for clustering. Further developments along these directions followed,
including work by \cite{shoham02,LLN2004,lin09,wang04,greselin10,andrews11b}.
An extension to mixture of $t$-factor analyzers (MtFA), adopting the family of MVT distributions
for the component factor and errors, was first considered by \cite{mclachlan07} and more recently
by \cite{andrews11a,andrews11c}.

In this paper, our objective is to illustrate the efficacy of a class of TMIX models with several
parsimonious covariance structures, called parsimonious $t$ mixture models (PTMMs).
The PTMMs are based on assuming a constrained $t$-factor structure on
each mixture component for parsimoniously modelling population heterogeneity in the presence of fat tails; the PTMMs
are $t$-analogues of the PGMMs. We are devoted to developing additional tools for a simplified version of the
MODt family of \cite{andrews11c} and applying the proposed tools on image reconstruction tasks.

The EM algorithm \citep{dempster77} and its extensions,
such as the expectation conditional maximization (ECM) algorithm \citep{meng93}
and the expectation conditional maximization either (ECME) algorithm \citep{liu94,liu95}, have been
practiced as useful tools for conducting  maximum likelihood (ML) estimation in a variety of mixture modelling scenarios.
To improve the computational efficiency for fitting PTMMs,
we adopt a three-cycle AECM algorithm
\citep{meng97} that allows specification of different complete data at each cycle.

The rest of the paper is organized as follows. In Section~\ref{sec:tfac},
we briefly describe the single $t$ factor analysis model and study some related properties.
In Section~\ref{sec:ptmm}, we present the formulation of PTMM and discuss the methods for fitting these models.
In Section~\ref{sec:apps}, we demonstrate how PTMM can be applied
to image compression and compact facial representation tasks.
Some concluding remarks are given in Section~\ref{sec:conc}.

\section{The $t$ factor analysis model}\label{sec:tfac}

We briefly review the $t$ factor analysis (tFA) model, which can be thought of as
a single component MtFA model. Let $\bmY=(\bmY_1,\ldots,\bmY_n)$ be a set of $p$-dimensional random vectors.
In the tFA setting, each $\bmY_j$ is modelled as
\begin{equation}
\bmY_j=\bmmu+\bmB\bmU_j+\bmvarepsilon_j
\label{tFAmodel}
\end{equation}
with
\begin{eqnarray*}
\left[\begin{array}{cc}\bmU_j\\\bmvarepsilon_j\end{array}\right]\sim
t_{q+p}\left(
\left[\begin{array}{cc}\bmzero\\\bmzero\\\end{array}\right],
\left[\begin{array}{cc}\bmI_q & \bmzero\\\bmzero & \bmPsi\\\end{array}\right],\nu\right),\end{eqnarray*}
where $\bmmu$ is a $p$-dimensional mean vector, $\bmB$ is a $p\times q$ matrix of factor loadings,
$\bmU_j$ is a $q$-dimensional ($q\ll p$) vector of latent variables called {\em common factors}, $\bmvarepsilon_j$ is $p$-dimensional vector of errors called {\em specific factors}, $\bmI_q$ is a $q$-dimensional identity matrix,
$\bmPsi={\rm diag}\{\psi_1,\ldots,\psi_p\}$ is a diagonal matrix with positive entries, and $\nu$ is the df.

Note that $\bmU_j$ and $\bmvarepsilon_j$ are uncorrelated and marginally $t$-distributed, but not independent.
Using the characterization of the $t$ distribution, model~(\ref{tFAmodel}) can be hierarchically represented as
\begin{eqnarray*}
\bmY_j\mid(\bmU_j,\tau_j)&\sim& N_p(\bmmu+\bmB\bmU_j,\tau_j^{-1}\bmPsi),\\
\bmU_j\mid\tau_j&\sim& N_q(\bmzero,\tau^{-1}_j\bmI_q),\\
\tau_j&\sim& {\rm Gamma}(\nu/2,\nu/2),\label{hier1}
\end{eqnarray*}
where ${\rm Gamma}(\alpha,\beta)$ stands for the gamma distribution with mean $\alpha/\beta$.
It follows that the marginal distribution of $\bmY_j$, after integrating out $\bmU_j$ and $\tau_j$, can
be expressed as $\bmY_j\sim t_{p}(\bmmu,\bmB\bmB^{\rm T}+\bmPsi,\nu)$.

When handling high-dimensional data for large $p$ relative to $n$, say $p\gg n$,
the inverse of $\bmB\bmB^{\rm T}+\bmPsi$ plays an important role in computational complexity. In such a case,
we use the following matrix inversion formula \citep{woodbury50}:
\BE
 (\bmB\bmB^{\rm T}+\bmPsi)^{-1}
 =\bmPsi^{-1}-\bmPsi^{-1}\bmB(\bmI_q+\bmB^{\rm T}\bmPsi^{-1}\bmB)^{-1}\bmB^{\rm T}\bmPsi^{-1},\label{Sinv}
\EE
which can be done more quickly because it involves only the low-dimensional $q\times q$ inverse plus the inversion of a diagonal $p\times p$ matrix.
Moreover, the determinant of $\bmB\bmB^{\rm T}+\bmPsi$ can be calculated as
\BE
  |\bmB\bmB^{\rm T}+\bmPsi| = |\bmI_q+\bmB^{\rm T}\bmPsi^{-1}\bmB| \prod_{i=1}^p \psi_{i}.\label{Det}
\EE
The formulae (\ref{Sinv}) and (\ref{Det}) have been used many times before, including work by \cite{mclachlan00a} and \cite{mcnicholas08}.

\section{Parsimonious multivariate $t$ mixture models}\label{sec:ptmm}

Consider $n$ independent $p$-dimensional random vectors $\bmY_1,\ldots,\bmY_n$ that come from a heterogeneous population with $g$ non-overlapping classes. Suppose that the density function of each feature $\bmY_j$
can be modelled by a $g$-component MtFA:
\begin{eqnarray}
f(\bm Y_j\mid\bmTheta)=\sum^g_{i=1}w_i t_p(\bm Y_j|\bmmu_i,\bmB_i\bmB^{\rm T}_i+\bmPsi_i, \nu_i),\label{mix.dist}
\end{eqnarray}
where the $w_i$ are mixing proportions that are constrained to be positive and
$\sum_{i=1}^gw_i=1$, $\bmB_i$ is a $p\times q$ matrix of factor loadings, and $\bmPsi_i={\rm diag}\{\psi_{i1},\ldots,\psi_{ip}\}$ is a
diagonal matrix. We use $\bmTheta=(w_1,\ldots,w_{g}, \bmtheta_1, \ldots,  \bm\theta_g)$ to represent all unknown parameters with
$\bmtheta_i=(\bmmu_i, \bmB_i, \bm\Psi_i, \nu_i)$ containing the parameters for the density of component $i$.

As in \cite{mclachlan07}, the MtFA model can be alternatively formulated by exploiting its link
with the tFA model. Specifically, we assume that
$$\bmY_j=\bmmu_i+\bmB_i\bmU_{ij}+\bmvarepsilon_{ij}$$
with probability~$w_i$, for $i=1,\ldots,g$ and $j=1,\ldots,n$, where $\bmU_{ij}$ and $\bmvarepsilon_{ij}$ are,
respectively, the common and specific factors corresponding to component $i$. We assume that $\bmU_{ij}$
and $\bmvarepsilon_{ij}$ have a joint MVT distribution to be consistent
with model (\ref{mix.dist}) for the marginal distribution of $\bmY_j$.
Under a hierarchical mixture modelling framework, each $\bmY_j$
is conceptualized to have originated from one of $g$ classes. It is convenient to construct unobserved
allocation variables $\bmZ_{j} = (Z_{1j},\ldots,Z_{gj})^{\rm T}$,  for $j=1,\ldots,n$, whose values are a set of binary
variables with $Z_{ij}=1$ indexing so that $\bmY_j$ belongs to class $i$ and are
constrained to be $\sum_{i=1}^g Z_{ij}=1$ for each $j$.
More specifically, $\bmZ_j$ is distributed as
a multinomial random vector with one trial and cell probabilities $w_1,\ldots,w_g$, denoted by $\bmZ_j\sim \MM(1; w_1,\ldots,w_g)$.
After a little algebra \citep[cf.][]{mclachlan07}, we have
\BE
\bmU_{ij}\mid (\bmY_j,Z_{ij}=1)
\sim t_q\big(\bmGamma^{\rm T}_i(\bmY_j-\bmmu_i),\frac{\nu+\delta_{ij}}{\nu+p}\,\bmOmega_i,\nu+p\big),\label{postUij}
\EE
where $\bmGamma_i=(\bmB_i\bmB^{\rm T}_i+\bmPsi_i)^{-1}\bmB_i$, $\bmOmega_i=\bmI_q-\bmGamma^{\rm T}_i\bmB_i$
and $\delta_{ij}=(\bmY_j-\bmmu_i)^{\rm T}(\bmB_i\bmB^{\rm T}_i+\bmPsi_i)^{-1}(\bmY_j-\bmmu_i)$ denotes
the Mahalanobis-squared distance between $\bmY_j$ and $\bmmu_i$.


Following \cite{mcnicholas08}, we extend the MtFA model by allowing
constraints across groups on matrices $\bmB_i$ and $\bmPsi_i$. Specifically, we allow the following constraints on
the scaling covariance matrices:
$\bmB_i=\bmB$; $\bmPsi_i=\bmPsi$; or $\bmPsi_i=\psi_i\bmI_p$, for $i=1,\ldots,g$.
In addition, we consider the
constraint $\nu_i=\nu~(i=1,\ldots,g)$ when we impose this constraint to reduce the
number of parameter in PTMMs, our family of models (Table~\ref{tab:ptmms}) is called `PTMM1', and when the $\nu_i$ are
allowed to vary across components, we call our family `PTMM2'.
In this latter case (PTMM2), one simply adds $g-1$ parameters to the last column of Table~\ref{tab:ptmms}.
The number of free covariance parameters in the PTMM family --- the union of PTMM1 and PTMM2 ---
can be reduced to as few as $qp-q(q-1)/2+2$ or inflated to as many as $g[pq-q(q-1)/2+p+1]$.
\begin{table*}[t]
\begin{center}
\caption{The eight covariance structures, used by \cite{mcnicholas08}, that form the
basis of the PTMM considered herein. The ``Constrained'' in the columns ``Loading'',
``Error'', and ``Isotropic'' represent $\bmB_i=\bmB$, $\bmPsi_i=\bmPsi$ and
$\psi_{ik}=\psi_i$ for $i=1,\ldots,g$ and $k=1,\ldots,k$, respectively.}\label{tab:ptmms}
\scriptsize
\begin{tabular*}{
1\textwidth}{@{\extracolsep{\fill}}|l|rrr|r|}\hline
Structure   & Loading  & Error & Isotropic &  Number of parameters\\  \hline
 CCC    &   Constrained & Constrained &   Constrained   &  $(pq-q(q-1)/2)+1+1$ \\
 CCU    &   Constrained & Constrained & Unconstrained   &  $(pq-q(q-1)/2)+p+1$ \\
 CUC    &   Constrained & Unconstrained &   Constrained   &  $(pq-q(q-1)/2)+g+1$ \\
 CUU    &   Constrained &Unconstrained& Unconstrained   &  $(pq-q(q-1)/2)+gp+1$\\
 UCC    & Unconstrained & Constrained  &   Constrained   &  $ g(pq-q(q-1)/2)+1+1$\\
 UCU    & Unconstrained & Constrained & Unconstrained   &  $ g(pq-q(q-1)/2)+p+1$\\
 UUC    & Unconstrained &Unconstrained&  Constrained     &  $ g(pq-q(q-1)/2)+g+1$\\
 UUU    & Unconstrained &Unconstrained& Unconstrained   &  $ g(pq-q(q-1)/2)+gp+1$\\ \hline
\end{tabular*}
\end{center}
\end{table*}

The implementation of an efficient 3-cycle AECM algorithm for estimating PTMMs together with
some practical issues including the specification of starting values, the stopping rule
and the model selection criterion are described in
Supplementary Material.

\section{Applications}\label{sec:apps}

\subsection{Image compression}
 A colour image is a digital image that includes 24-bit RGB colour information for each pixel.
An RGB image is derived from the three primary colours --- red~(R), green~(G), and blue~(B) ---
for which each colour is 8 bits. The technique of image compression plays a central role in image
transmission and storage for high quality. There are several unsupervised image compression
approaches based on probabilistic models. The MFA model is well recognized as a dominant dimension
reduction technique and is useful in block image transform coding \citep{ueda00}. In this subsection,
we individually apply the PTMMs and PGMM to colour image compression and compare the quality of the
reconstructed images. A $512\times 512$ RGB colour image `Lena' (encoded in 24 bits per pixel) is
subdivided into $n=16384$ non-overlapping RGB-blocks of $4 \times 4 \times 3$ pixels
and each block is taken as a 48-dimensional data vector $\bmy$. Let $\mathcal{Y}$ be a set of collected $\bmy$ vector.
Making a slight modification of \cite{ueda00}, our compression procedure to transform the experimental
image is summarized below.
\begin{description}
\item[1.] Set the desired number of components $g$ and dimensionality of factors $q$.
\item[2.] Estimate $\bmTheta$ by fitting a parsimonious mixture model to $\mathcal{Y}$.
\item[3.] Perform a model-based clustering according to the component membership of data point $\bmy$, which is
decided by maximizing the posterior probability $\Pr(Z_{ij}=1|\bmy)$.
\item[4.] For each $\bmy_j\in\mathcal{Y}$ classified to $\mathcal{C}_{i}$, calculate
\BE
  \hat{\bmy}_j=\hat{\bmmu}_{i}+\hat{\bmB}_{i}\hat{\bmU}_{ij},\label{hatyj}
\EE
where $\hat{\bmU}_{ij}=\hat{\bmB}^{\rm T}_i(\hat{\bmB}_i\hat{\bmB}^{\rm T}_i+\hat{\bmPsi}_i)^{-1}(\bmY_j-\hat{\bmmu}_i)$ is
the estimated posterior mean of (\ref{postUij}).
\item[5.] Reconstruct $\hat{\bmy}_j$, $j=1,\ldots,n$, into a compressed color image.
\end{description}

In what follows,
we assumed that the characteristic of each components in the mixture model are the same, which leads to the
MFA model with common factor loadings (CUU structure).
The Lena image data are fitted by the CUU structure with $(g=4,q=4)$ and $(g=8,q=8)$ using the PGMM and PTMM approaches.
To evaluate the quality of the reconstructed image, we compute the root mean squared error (RMSE) and
peak signal-to-noise ratio (PSNR), which are widely used metrics in the image coding field. For an original
color image $\{I_R(x,y),I_G(x,y),I_B(x,y)\}$ of size $M\times N$ and a reconstructed
image $\{\hat{I}_R(x,y),\hat{I}_G(x,y),\hat{I}_B(x,y)\}$, the RMSE is defined as
\[
  \mbox{RMSE}=\sqrt{\frac{1}{3}\left(MSE_{R}+MSE_{G}+MSE_{B}\right)},
\]
where
\[
  MSE_{R}=\frac{1}{MN}\sum_{x=1}^M\sum_{y=1}^N\big[I_R(x,y)-\hat{I}_R(x,y)\big]^2,
\]
and the other two quantities $MSE_{G}$ and $MSE_{B}$ are defined in the same fashion.
The measure RMSE represents the average Euclidean distance from the original image to the output one.
For a 24-bit colour $[0,255]^3$ image, the PSNR is given by
\[
  \mbox{PSNR}=20\log_{\rm 10}\left(\frac{255}{\mbox{RMSE}}\right).
\]
Note that the higher PSNR value indicates a compressed (i.e., reconstructed) image of better quality.

Experimental results are provided in Table~\ref{tab2}, including the log-likelihoods, the number of parameters,
and BIC values, together with the RMSE and PSNR for the compressed images. The performance of PTMM1 and
PTMM2 are almost the same in terms of RMSE and PSNR for the compressed images. Notably, the BIC and PSNR
values obtained from the PTMM approaches are much higher than the PGMM for all cases.
\begin{table}[t]
\scriptsize
\begin{center}
\caption{Comparisons of fitting adequacy and quality of image reconstruction using three CUU models with $(g=4;q=4)$
and $(g=8;q=8)$.}
\begin{tabular*}{
.8\textwidth}{@{\extracolsep{\fill}}|lrrrrr|}\hline
  \multicolumn{6}{|c|}{$(g=4;~q=4)$}\\
  \hline
   Model     &  $\ell_{\mbox{max}}(\times 10^3)$ & $m$ & BIC$(\times 10^3)$  & RMSE  &  PSNR\\  \hline
 PGMM      & 2104.4 & 189 & 4206.8 & 7.62 & 30.5\\
 PTMM1     & 2205.9 & 190 & 4409.7 & 5.22 & 33.8\\
 PTMM2     & 2204.6 & 193 & 4407.1 & 5.26 &33.7\\
 \hline
 \multicolumn{6}{|c|}{$(g=8;~q=8)$}\\
  \hline
   Model     &  $\ell_{\mbox{max}}(\times 10^3)$ & $m$ & BIC$(\times 10^3)$  & RMSE  &  PSNR\\  \hline
 PGMM      & 2235.1&363&4466.3& 7.03 &  31.2\\
 PTMM1     & 2348.8&364&4693.7& 2.78 &  39.2\\
 PTMM2     & 2351.2&371&4698.5& 2.43 & 40.4\\
 \hline
\end{tabular*}\label{tab2}
\end{center}
\hspace*{1.7cm}$\mbox{BIC}=2\ell_{\rm max}-m\log n$. Models
with large BIC scores are preferred.
\end{table}

\begin{figure}[h]
\centering
\epsfig{file=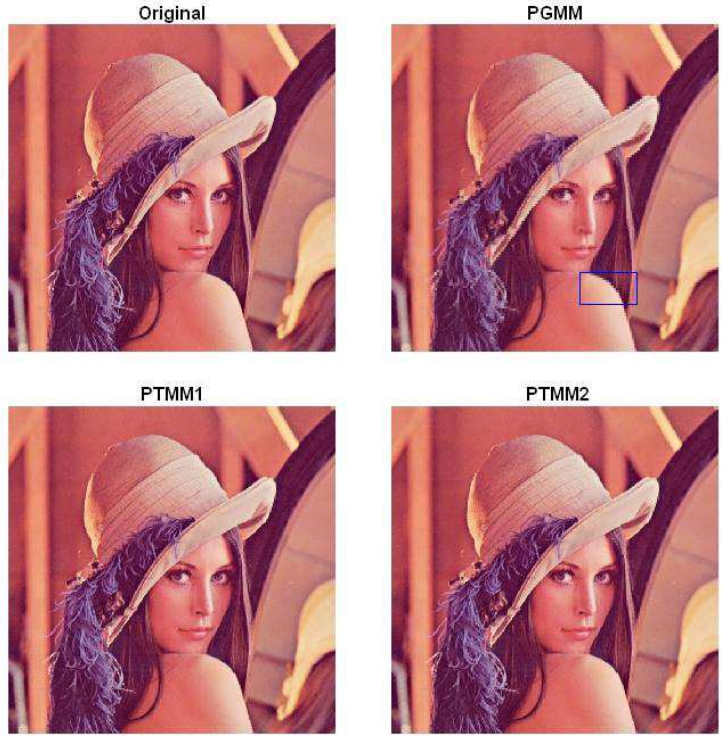, width=4in}
\caption{An example of Lena colour image reconstruction using three CUU models with $g=8$ and $q=8$.}\label{fig1}
\end{figure}

Figure~\ref{fig1} shows the original Lena image and three compressed images obtained by fitted the CUU $(g=8; q=8)$ models. The quality of the reconstructed images using the PTMMs
 is much better (i.e., smoother) than those reconstructed using the PGMM (for instance, observe the edge of Lena's shoulder).

\subsection{Compact facial representation}
Over the past few decades, numerous template matching approaches, along with their modifications and variants,
have been proposed for human perception system face recognition; see \cite{zhao03} for an extensive survey.
Motivated by a technique developed by \cite{sirovitch87}, the eigenface method \citep{turk91} has become the
most popular tool for facial feature extraction in computer vision research. Its implementation is primarily based
on the use of PCA, also known as the Karhunen-Lo\`{e}ve transform. The central idea of
PCA is to find a multilinear subspace whose orthonormal basis maximizes the scatter matrix of all the projected samples.

Although the eigenface method is remarkably simple and quite straightforward,
its performance depends heavily on pose, lighting and background conditions.
Because PCA is a one-sample method, a single set of eigenfaces may not be enough to
represent face images having a large amount of variation.
Some efforts have been made to generalize it to the mixture-of-eigenfaces method \cite{kim02}, which generates more than one set of eigenfaces for better representation of the whole face.
However, such a method will produce an over-parameterized solution in many circumstances.
To capture the heterogeneity among the underlying face images while maintaining parsimony in modeling,
we apply the PTMM approach with common factor loadings (CUU) as a novel strategy for robust facial representation.

Consider a training set of $n$ face images ${I_1,\ldots,I_n}$. Suppose that all images are exactly the same size; $a$ pixels in width by~$b$ in height.
Let $\bmy=\{\bmy_j={\rm vec}(I_j)\}_{j=1}^n$ be the transformed image vectors taking intensity values in a $p$-dimensional
image space, where $p=a\times b$. In what follows, we briefly review the eigenface method. Define the total scatter matrix of the sample images as
\[
  \bmS_T=\sum_{j=1}^n(\bmy_j-\bar{\bmy})(\bmy_j-\bar{\bmy})^{\rm T},
\]
where  $\bar{\bmy}$ is the mean image vector of all samples.
In PCA, the optimal orthonormal basis vectors $\bmOmega_{\rm PCA}\in R^{p\times K}$,
where $K<<p$, are chosen to maximize the following objective function
\[
   \bmE_{\rm PCA}=[\bme_1\cdots \bme_{K}]=\mbox{argmax}_{\bmE}|\bmE^{\rm T}\bmS_T\bmE|,
\]
where $\{\bme_k\}_{k=1}^K$ is the set of $p$-dimensional vectors of $\bmS_T$ corresponding to the $K$ largest eigenvectors,
which are referred to as eigenfaces in \cite{turk91} because they have exactly the same dimension as the original images.
Each face image vector $\bmy_i$ can be represented as a linear combination of the best $K$ eigenvectors:
\BE
      \hat{\bmy}^{{\rm PCA}}_j= \bar{\bmy}+\sum_{k=1}^K\omega_{jk}\bme_k,\label{yhat-pca}
\EE
where $\omega_{jk}=\bme^{\rm T}_k (\bmy_j-\bar{\bmy})$.

As another illustration, we compare the face reconstruction performance among the PGMM and PTMM approaches using
the image compression algorithm described in the above example and the eigenface method.
We use the face images in the Yale face database, which contains 165 grayscale images of 15 individuals in GIF format.
There are 11 images per person with pose and lighting variations, but for illustrative purposes and
simplicity we consider only 11 images of one individual. For each image, we manually select the centers of eyes, denoted by $(x,y)$.
The four boundaries of each cropped central part of the face are $x-60$, $x+60$, $y-100$, and $y+60$, which gives
images of $121\times 161$ pixels as a sample in our experiment.

In this experiment, the 11 face images are trained by seven models: the PCA, PGMM, and PTMM1 models with CUU structure, where $K=q=5$ and $g=1,2,3$.
The reconstructed images can be obtained by following two steps.
\begin{description}
  \item[1.] In each model, the fitted vector is obtained as a reconstructed image, say $\hat\bmy_j$. For the PCA method,
  $\hat\bmy_j=\hat\bmy_j^{{\rm PCA}}$ as defined in (\ref{yhat-pca}).
 As for the PGMM and PTMM1, the solution is obtained by using (\ref{hatyj}).
  \item[2.] If the entries of $\hat\bmy_j$ do not fall in the color space, such as $[0,1]$, the reproductions $\hat\bmy_j^\ast$ are normalized by
    \[\hat y_{jr}^\ast=\frac{\hat y_{jr}-\min\{\hat\bmy_j\}}{\max\{\hat\bmy_j\}-\min\{\hat\bmy_j\}},\quad\hbox{for $r=1,\ldots,p$}.\]
\end{description}

\begin{figure}[!t]
\centering
\epsfig{file=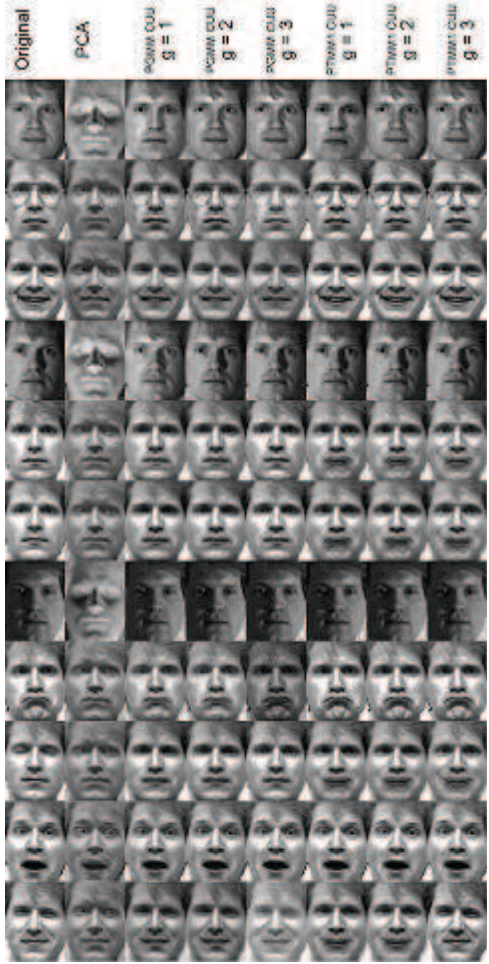, scale=1, angle=0}
\caption{Reconstructed images based on PCA, PGMM and PTMM approaches. The first column shows sample images for one
individual of the Yale database.}\label{reconst}
\end{figure}

\begin{figure}[!t]
\centering
\epsfig{file=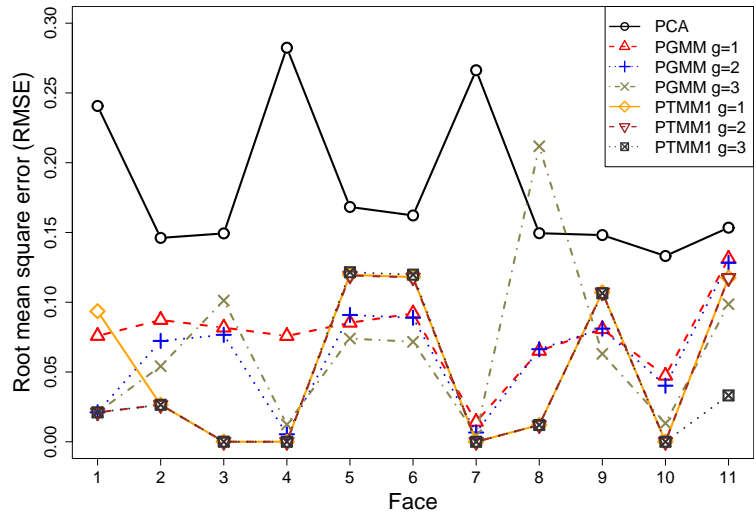, scale=.35, angle=-90}
\caption{RMSE values of seven trained models for 11 face images.}\label{rmse_traj}
\end{figure}

Figure \ref{reconst} shows the original and the reconstructed images obtained from the seven training models. PCA clearly has the worst
performance, while the PGMM and PTMM lead to somewhat comparable results. For explicitly measuring the reconstruction quality,
we further calculated the RMSE, denoted by
$[(\bmy_j-\hat\bmy_j^\ast)^\top(\bmy_j-\hat\bmy_j^\ast)]^{1/2}$, for each image and each model. Figure \ref{rmse_traj} displays the patterns of RMSE values. When comparing these models, smaller RMSEs indicate better reconstructions. In general, PTMM1 yields lower RMSE values than those from PGMM. As a result, we conclude that PTMM1 can be a prominent tool for facial coding.

\clearpage

\section{Concluding remarks}\label{sec:conc}
We have utilized a class of parsimonious $t$ mixture models, called the PTMMs,
which may create tremendous flexibility in robust clustering of high-dimensional data as they are relatively insensitive to outliers.
This model-based tool allows practitioners to analyze heterogeneous
multivariate data in a broad variety of considerations and works particularly well in high-dimensional settings.
Numerical results show that the proposed PTMM approach performs reasonably well for the experimental image data.
As pointed by by \cite{ZhaoYu2008} and \cite{WangLin2013}, the convergence of the AECM algorithm can be painfully slow in certain
situations. It is a  worthwhile task to pursue some modified algorithms toward fast convergence.

\bibliographystyle{kp}

\bibliography{cosbib}

\newpage
\section*{Appendix}
\section*{A: ML estimation via the AECM algorithm}
\hspace*{.6cm}
We discuss how to carry out the AECM algorithm for computing the ML estimates of the parameters in PTMMs.
To formulate the algorithm which consists of three cycles,
we partition the unknown parameters $\bmTheta$ as $(\bmTheta_1,\bmTheta_2,\bmTheta_3)$, where $\bmTheta_1$ contains $w_i$'s, $\bmTheta_2$ contains
$\bmmu_i$'s and $\nu_i$'s,
while $\bmTheta_3$ contains $\bmB_i$'s and $\bmPsi_i$'s. Let $\bmZ=(\bmZ_1, \ldots, \bmZ_n)$ be the collection of
latent group labels.
In the first cycle of the algorithm, we treat $\bmZ$ as missing data. So the complete data is
$\Yaug=(\bmY,\bmZ)$ and our aim is to estimate
$\bmTheta_1$ with $\bmTheta_2$ and $\bmTheta_3$ fixed at their current estimates $\hat{\bmTheta}_2$ and $\hat{\bmTheta}_3$.\\
\hspace*{.6cm}
The log-likelihood of $\bm\Theta_1$ based on the complete data $\Yaug$, apart from an additive constant, is given by
\begin{eqnarray}
          \ell_c^{[1]}(\bmTheta_1|\Yaug)&=&
          \sum_{j=1}^n\sum_{i=1}^gZ_{ij}\log w_i.\label{mix.ellc1}
        \end{eqnarray}
Conditioning on $\bmY$ and $\hat{\bmTheta}$, taking expectation for (\ref{mix.ellc1}) leads to
\begin{eqnarray}
          Q^{[1]}(\bmTheta_1|\hat{\bm\Theta})&=&
          \sum_{i=1}^g\sum_{j=1}^n\hat{z}_{ij}\log w_i\label{mix.Q1}
        \end{eqnarray}
with
    \[\hat{z}_{ij}=\frac{\hat{w}_i\,t_p(\bm Y_{j}\mid\hat{\bmmu}_i,\hat{\bmB}_i\hat{\bmB}^{\rm T}_i+\hat{\bmPsi}_i,\hat{\nu}_i)}
{\sum_{h=1}^g\hat{w}_h\,t_p(\bm Y_{j}\mid\hat{\bmmu}_h,\hat{\bmB}_h\hat{\bmB}^{\rm T}_h+\hat{\bmPsi}_h,\hat{\nu}_h)}.\]

Maximizing (\ref{mix.Q1}) with respect to $w_i$, that restricted to $\sum_{i=1}^gw_i=1$, yields
\[
\hat{w}_i=\frac{\sum_{j=1}^n\hat{z}_{ij}}{n}.
\]
\hspace*{.6cm}
At the second cycle, when estimating $\bmTheta_2$, we treat $\bmtau$ and $\bmZ$ as the missing data.
The log-likelihood function of $\bmTheta_2$ based
on the complete data $\Yaugtwo=(\Yaug,\bmtau)$ takes the form of
\begin{eqnarray}
          \ell_c^{[2]}(\bmTheta_2|\Yaugtwo)&=&
          \sum_{j=1}^n\sum_{i=1}^gZ_{ij}\Big\{-\frac12\tau_{j}(\bm Y_{j}-\bmmu_i)^{\rm T}(\hat{\bmB}_i\hat{\bmB}_i^{\rm T}+\hat{\bmPsi}_i)^{-1}(\bm Y_{j}-\bmmu_i)\nonumber\\
                                      &&+\frac{\nu_i}{2}\log\Big(\frac{\nu_i}{2}\Big)-\log\Gamma\Big(\frac{\nu_i}{2}\Big)+\frac{\nu_i}{2}(\log\tau_{j}-\tau_{j})\Big\}. \label{mix.ellc2}
        \end{eqnarray}
Conditioning on $\bmY$ and $\hat{\bmTheta}$, taking expectation for (\ref{mix.ellc2}) leads to
\begin{eqnarray}
          Q^{[2]}(\bmTheta_2|\hat{\bm\Theta})&=&
          -\frac12\sum_{i=1}^g\sum_{j=1}^n\hat{z}_{ij}\hat{\tau}_{ij}(\bmY_{j}-\bmmu_i)^{\rm T}(\hat{\bmB}_i\hat{\bmB}_i^{\rm T}+\hat{\bmPsi}_i)^{-1}(\bm Y_{j}-\bmmu_i)\nonumber\\
&&+\sum_{i=1}^g\sum_{j=1}^n\hat{z}_{ij}\Big[\frac{\nu_i}{2}\log\Big(\frac{\nu_i}{2}\Big)-\log\Gamma\Big(\frac{\nu_i}{2}\Big)+\frac{\nu_i}{2}(\hat{\kappa}_{ij}-\hat{\tau}_{ij})\Big]\label{mix.Q2}
        \end{eqnarray}
with
\begin{eqnarray*}
\hat{\tau}_{ij}&=&\frac{\hat{\nu}_i+p}{\hat{\nu}_i+\hat{\delta}_{ij}}~~\mbox{and}~~\hat{\kappa}_{ij}=\mbox{DG}\left(\frac{\hat{\nu}_i+p}{2}\right)-\log\left(\frac{\hat{\nu}_i+\hat{\delta}_{ij}}{2}\right),
\end{eqnarray*}
where
$\hat{\delta}_{ij}=(\bmY_{j}-\hat{\bmmu}_i)^{\rm T}(\hat{\bmB_i}\hat{\bmB_i}^{\rm T}+\hat{\bmPsi}_i)^{-1}(\bmY_{j}-\hat{\bmmu}_i)$ and DG$(\alpha)=\Gamma'(\alpha)/\Gamma(\alpha)$ is the digamma function.\\
\hspace*{.6cm}
Maximizing (\ref{mix.Q2}) with respect to $\bmmu_i$ and $\nu_i$ yields
\[
\hat{\bmmu}_i=\frac{\sum_{j=1}^n\hat{z}_{ij}\hat{\tau}_{ij}\bm Y_{j}}{\sum_{j=1}^n\hat{z}_{ij}\hat{\tau}_{ij}},
\]
and
\[
\hat{\nu}_i={\rm argmax}_{\nu_i}\left\{\frac{\nu_i}{2}\log\Big(\frac{\nu_i}{2}\Big)-\log\Gamma\Big(\frac{\nu_i}{2}\Big)+\frac{\nu_i}{2}
\left[\frac{\sum^n_{j=1}\hat{z}_{ij}(\hat\kappa_{ij}-\hat\tau_{ij})}{\sum^n_{j=1}\hat{z}_{ij}}\right]\right\},
\]
which is equivalently the solution of the following equation:
\[
\log\Big(\frac{\nu_i}{2}\Big)-{\rm DG}\Big(\frac{\nu_i}{2}\Big)+1+\frac{\sum^n_{j=1}\hat{z}_{ij}(\hat\kappa_{ij}-\hat\tau_{ij})}{\sum^n_{j=1}\hat{z}_{ij}}
      =0.
\]
In the case of $\nu_1=\cdots=\nu_g=\nu$, we obtain $\hat{\nu}$ as the solution of the following equation:
\[
\log\Big(\frac{\nu}{2}\Big)-{\rm DG}\Big(\frac{\nu}{2}\Big)+1+\frac{\sum^g_{i=1}\sum^n_{j=1}\hat{z}_{ij}(\hat\kappa_{ij}-\hat\tau_{ij})}{n}=0.
\]

At the third cycle, when estimating $\bmTheta_3$, we treat $\bmU$, $\bmtau$ and $\bmZ$ as the missing data.
The log-likelihood function of $\bmTheta_3$ based
on the complete data $\Yaugthree=(\Yaugtwo,\bmU)$ takes the form of
\begin{eqnarray}
 \ell_c^{[3]}(\bmTheta_3\mid\Yaugthree)&=&\sum_{i=1}^g\sum_{j=1}^nZ_{ij}\Big\{-\frac 12\log|\bmPsi_i|-
 \frac12\tau_{j}(\bm Y_{j}-\hat{\bmmu}_i)^{\rm T}\bmPsi_i^{-1}(\bm Y_{j}-\hat{\bmmu}_i)\nonumber\\
                              &&+\frac12\tau_{j}(\bm Y_{j}-\hat{\bmmu}_i)^{\rm T}\bmPsi_i^{-1}\bmB_i\bmU_{ij}
                              +\frac12\tau_{j}\bmU_{ij}^{\rm T}\bmB^{\rm T}_i\bm\Psi_i^{-1}(\bm Y_{j}-\hat{\bmmu}_i)\nonumber\\
                              &&-\frac12\tau_{j}\bmU_{ij}^{\rm T}\bmB^{\rm T}_i\bm\Psi_i^{-1}\bmB_i\bmU_{ij}\Big\}.\label{mix.ellc3}
\end{eqnarray}
Therefore, the expectation of (\ref{mix.ellc3}) conditioning on the observed data $\bmY$ and
the updated vales of $\hat{\bmTheta}$ is
\begin{eqnarray*}
Q^{[3]}(\bmTheta_3\mid\hat{\bm\Theta})&=&
\frac{1}{2}\sum_{i=1}^g\hat{n}_i\log|\bmPsi_i^{-1}|-\frac{1}{2}\sum_{i=1}^g\hat{n}_i
\mbox{tr}\Big\{\bmPsi_i^{-1}\bmS_i-\bm\Psi_i^{-1}\bmB_i\hat{\bmGamma}_i^{\rm T}\bmS_i\\\
&&-
\bmPsi_i^{-1}\bmS_i\hat{\bmGamma}_i\bmB_i^{\rm T}+\bmPsi_i^{-1}\bmB_i(\hat{\bm\Omega}_i+\hat{\bmGamma}_i\bmS_i\hat{\bmGamma}_i^{\rm T})\bmB_i^{\rm T}\Big\},
\end{eqnarray*}
where $\hat{n}_i=\sum_{j=1}^n\hat{z}_{ij}$, $\bmS_i=
\hat{n}^{-1}_i\sum_{j=1}^n\hat{z}_{ij}\hat{\tau}_{ij}(\bm  Y_{j}-\hat{\bmmu}_i)(\bm  Y_{j}-\hat{\bmmu}_i)^{\rm T}$ and
$\hat{\bm\Omega}_i=\bmI_q-\hat{\bm\Gamma}^{\rm T}_i\hat{\bmB}_i$.
The resulting CM-steps for the updates of $\bmB_i$ and $\bmPsi_i$ under eight
constrained/unconstrained situations are given in the next Section. Note that the above procedure
is also applicable for the tFA model by treating $Z_{ij}=1$.

\section*{B: Calculations for eight parsimonious models}
To facilitate the derivation, we adopt the following notation:
\[
\tilde{\bmS}=\sum_{i=1}^g \hat{w}_i\bmS_i=n^{-1}\sum_{i=1}^g\sum_{j=1}^n\hat{z}_{ij}\hat{\tau}_{ij}(\bm  Y_{j}-\hat{\bmmu}_i)(\bm  Y_{j}-\hat{\bmmu}_i)^{\rm T}
\]
\subsection*{B.1: Model CCC}
For model CCC, we have $\bmB_i=\bmB$, $\bm\Psi_i=\bm\Psi=\psi\bmI_p$.
 The $Q^{[2]}$-function can be expressed as
\[
Q^{[2]}
=\frac{np}{2}p\log\psi^{-1}-\psi^{-1}\frac{n}{2}\Big[\mbox{tr}\{\tilde{\bmS}\}-2\mbox{tr}\{\bmB\hat{\bm\Gamma}^{\rm T}\tilde{\bmS}\}
+\mbox{tr}\{\bmB(\hat{\bmOmega}+\hat{\bmGamma}^{\rm T}\tilde{\bmS}\hat{\bm\Gamma})\bmB^{\rm T}\}\Big],
\]
where $\hat{\bmGamma}=(\hat{\bmB}\hat{\bmB}^{\rm T}+\hat{\psi}\bmI_p)^{-1}\hat{\bmB}$ and $\hat{\bmOmega}=\bmI_q-\hat{\bmB}^{\rm T}\hat{\bmGamma}$.
Differentiating $Q^{[2]}(\bm\Theta_2|\hat{\bm\Theta})$ with
respect to $\bmB$ and $\psi^{-1}$, respectively, and set the derivations equal to zero, we obtain
\[
\hat{\bmB}=\tilde{\bmS}\hat{\bm\Gamma}(\tilde{\bm \Omega}+\hat{\bm\Gamma}^{\rm T}\tilde{\bmS}\hat{\bm\Gamma})^{-1}~~\mbox{and}~~
\hat{\psi}=\frac1p\mbox{tr}\{\tilde{\bmS}-\hat{\bmB}\hat{\bmGamma}^{\rm T}\tilde{\bmS}\}.
\]
\subsection*{B.2: Model CCU}
For model CCU, we have $\bmB_i=\bmB$, $\bm\Psi_i=\bm\Psi$. The $Q^{[2]}$-function can be expressed as
\[
Q^{[2]}(\bm\Theta_2|\hat{\bm\Theta})=\frac{n}{2}\log|\bm\Psi^{-1}|-\frac{n}{2}\Big[\mbox{tr}\{\bm\Psi^{-1}\tilde{\bmS}\}-2\mbox{tr}\{\bm\Psi^{-1}\bmB\hat{\bm\Gamma}^{\rm T} \tilde{\bmS}\}+\mbox{tr}\{\bm\Psi^{-1}\bmB(\tilde{\bm \Omega}+\hat{\bm\Gamma}^{\rm T}\tilde{\bmS}\hat{\bm\Gamma})\bmB^{\rm T}\}\Big],
\]
where $\hat{\bmGamma}=(\hat{\bmB}\hat{\bmB}^{\rm T}+\hat{\bm\Psi})^{-1}\hat{\bmB}$.
Differentiating $Q^{[2]}(\bm\Theta_2|\hat{\bm\Theta})$ with respect to $\bmB$ and $\bm\Psi^{-1}$, respectively, and set the results equal to zero, we obtain
\[
\hat{\bmB}=\tilde{\bmS}\hat{\bm\Gamma}(\tilde{\bmOmega}+\hat{\bmGamma}^{\rm T}\tilde{\bmS}\hat{\bmGamma})^{-1}~~\mbox{and}~~
\hat{\bmPsi}=\mbox{Diag}\{\tilde{\bmS}-\hat{\bmB}\hat{\bm\Gamma}^{\rm T}\tilde{\bmS}\}.
\]
\subsection*{B.3: Model CUC}
For model CUC, we have $\bmB_i=\bmB$, $\bm\Psi_i=\psi_i\bmI_p$. The $Q^{[2]}$-function can be expressed as
\[
Q^{[2]}=\sum^{g}_{i=1}\frac{\hat{n}_i}{2}\Big[p\log|\psi^{-1}_i|-\psi^{-1}_i\mbox{tr}\{\bmS_i\}+2\psi^{-1}_i\mbox{tr}\{\bmB\hat{\bm\Gamma}_i^{\rm T}\bmS_i\}
                          -\psi^{-1}_i\mbox{tr}\{\bmB(\hat{\bm\Omega}_i+\hat{\bm\Gamma}_i^{\rm T}\bmS_i\hat{\bm\Gamma}_i)\bmB^{\rm T}\}\Big],
\]
where $\hat{\bm \Gamma}_i=(\hat{\bmB}\hat{\bmB}^{\rm T}+\hat{\psi}_i\bmI_p)^{-1}\hat{\bmB}$ and $\hat{\bm\Omega}_i=\bmI_q-\hat{\bm\Gamma}_i^{\rm T}\hat{\bmB}$.
Differentiating $Q^{[2]}(\bm\Theta_2|\hat{\bm\Theta})$ with respect to $\bmB$ and $\psi_i^{-1}$,
respectively, and set the results equal to zero, we obtain
\[
\hat{\bmB}=\left[\sum_{i=1}^g\frac{\hat n_i}{\psi_i}\bmS_i\hat{\bm\Gamma}_i\right]\left[\sum_{i=1}^g\frac{\hat n_i}{\psi_i}(\hat{\bm\Omega}_i+\hat{\bm\Gamma}_i^{\rm T}\bmS_i\hat{\bm\Gamma}_i)\right]^{-1}
\]
and
\[
\hat{\psi}_i=\frac 1p\mbox{tr}\{\bmS_i-2\hat{\bmB}\hat{\bm\Gamma}_i^{\rm T}\bmS_i+\hat{\bmB}(\hat{\bm\Omega}_i+\hat{\bm\Gamma}_i^{\rm T}\bmS_i\hat{\bm\Gamma}_i)\hat{\bmB}^{\rm T}\}.
\]
\subsection*{B.4: Model CUU}
For model CUU we have $\bmB_i=\bmB$. The $Q^{[2]}$-function can be expressed as
\[
Q^{[2]}=\sum^{g}_{i=1}\frac{\hat{n}_i}{2}\Big[\log|\bmPsi^{-1}_i|-\mbox{tr}\{\bmPsi^{-1}_i\bmS_i\}+2\mbox{tr}\{\bmPsi^{-1}_i\bm B\hat{\bm\Gamma}_i^{\rm T}\bmS_i\}
                          -\mbox{tr}\{\bmPsi^{-1}_i\bm B(\hat{\bm\Omega}_i+\hat{\bm\Gamma}_i^{\rm T}\bmS_i\hat{\bm\Gamma}_i)\bm B^{\rm T}\}\Big],
\]
where $\hat{\bm \Gamma}_i=(\hat{\bmB}\hat{\bmB}^{\rm T}+\hat{\bmPsi}_i)^{-1}\hat{\bmB}$ and $\hat{\bm\Omega}_i=\bmI_q-\hat{\bm\Gamma}_i^{\rm T}\hat{\bmB}$.
Differentiating $Q^{[2]}(\bm\Theta_2|\hat{\bm\Theta})$ with respect to $\bmB$ and $\bmPsi_i^{-1}$, respectively, and set the results equal to zero, we have
\[
\hat{\bmPsi}_i=\mbox{diag}\{\bmS_i-2\hat{\bmB}\hat{\bm\Gamma}_i^{\rm T}\bmS_i+\hat{\bmB}(\hat{\bm\Omega}_i+\hat{\bm\Gamma}_i^{\rm T}\bmS_i\hat{\bm\Gamma}_i)\hat{\bmB}^{\rm T}\}.
\]
However, there is no closed form for the loading matrix $\hat{\bmB}$. One can solve it through a row-by-row manner. Set
\[
\hat{\bmB}_{(p\times q)}=\left[ \hat{\bmb}^{\rm T}_1  \cdots  \hat{\bmb}^{\rm T}_k  \cdots  \hat{\bmb}^{\rm T}_p  \right]^{\rm T},
\]
where $\hat{\bmb}_k=[\hat{b}_{k1}~\hat{b}_{k2}~\cdots~\hat{b}_{kq}]$ represents the $k$th row of the matrix $\hat{\bmB}$.
Let $\bmR=\sum_{i=1}^g\hat n_i\bmPsi_i^{-1}\bmS_i\hat{\bm\Gamma}_i$ and $\bm r_h$ be $h$th row of the matrix $\bmR$. The $h$th
row of $\hat{\bmB}$ can be expressed as
\[
\hat{\bm b}_h\sum_{i=1}^g\frac{\hat n_i}{\hat{\psi}_{i(h)}}(\hat{\bm\Omega}_i+\hat{\bm\Gamma}_i^{\rm T}\bmS_i\hat{\bm\Gamma}_i)=\bm r_h,
\]
where $\psi_{i(h)}$ denotes the $h$th element along the diagonal of $\bmPsi_i$. Hence,
\[
\hat{\bm b}_h=\bm r_h\left[\sum_{i=1}^g\frac{\hat n_i}{\hat{\psi}_{i(h)}}(\hat{\bmOmega}_i+\hat{\bmGamma}_i^{\rm T}\bmS_i\hat{\bmGamma}_i)\right]^{-1},~~\mbox{for}~h=1, \ldots, p.
\]
\subsection*{B.5: Model UCC}
For model UCC we have $\bmPsi_i=\psi\bmI_p$. The $Q^{[2]}$-function can be expressed as
\[
Q^{[2]}=\sum^{g}_{i=1}\frac{\hat{n}_i}{2}\Big[p\log|\psi^{-1}|-\psi^{-1}\mbox{tr}\{\bmS_i\}+2\psi^{-1}\mbox{tr}\{\bmB_i\hat{\bm\Gamma}_i^{\rm T}\bmS_i\}
                          -\psi^{-1}\mbox{tr}\{\bmB_i(\hat{\bm\Omega}_i+\hat{\bm\Gamma}_i^{\rm T}\bmS_i\hat{\bm\Gamma}_i)\bmB_i^{\rm T}\}\Big],
\]
where $\hat{\bmGamma}_i=(\hat{\bmB_i}\hat{\bmB}_i^{\rm T}+\hat{\psi}\bmI_p)^{-1}\hat{\bmB}_i$ and $\hat{\bm\Omega}_i=\bm I_q-\hat{\bm\Gamma}_i^{\rm T}\hat{\bmB}_i$.
Differentiating $Q^{[2]}(\bm\Theta_2|\hat{\bm\Theta})$ with respect to $\bmB_i$ and $\psi^{-1}$, respectively, and set the results equal to zero, we obtain
\[
\hat{\bmB}_i=\bmS_i\hat{\bmGamma}_i(\hat{\bmOmega}_i+\hat{\bmGamma}_i^{\rm T}\bmS_i\hat{\bmGamma}_i)^{-1}
~~\mbox{and}~~
\hat{\psi}=\frac1p\sum_{i=1}^g\hat w_i\mbox{tr}\{\bmS_i-\hat{\bmB}_i\hat{\bm\Gamma}_i^{\rm T}\bmS_i\}.
\]
\subsection*{B.6: Model UCU}
For model UCU we have $\bmPsi_i=\bmPsi$. The $Q^{[2]}$-function can be expressed as
\[
Q^{[2]}=\sum^{g}_{i=1}\frac{\hat{n}_i}{2}\Big[\log|\bmPsi^{-1}|
-\mbox{tr}\{\bmPsi^{-1}\bmS_i\}+2\mbox{tr}\{\bmPsi^{-1}\bmB_i\hat{\bm\Gamma}_i^{\rm T}\bmS_i\}
-\mbox{tr}\{\bmPsi^{-1}\bmB_i(\hat{\bmOmega}_i+\hat{\bmGamma}_i^{\rm T}\bmS_i\hat{\bmGamma}_i)\bmB_i^{\rm T}\}\Big],
\]
where $\hat{\bmGamma}_i=(\hat{\bmB_i}\hat{\bmB}_i^{\rm T}+\hat{\bmPsi})^{-1}\hat{\bmB}_i$ and $\hat{\bmOmega}_i
=\bmI_q-\hat{\bmGamma}_i^{\rm T}\hat{\bmB}_i$.
Differentiating $Q^{[2]}(\bm\Theta_2|\hat{\bmTheta})$ with respect to $\bmB_i$ and $\bmPsi^{-1}$, respectively,
and set the results equal to zero, we obtain
\[\hat{\bmB}_i=\bmS_i\hat{\bm\Gamma}_i(\hat{\bmOmega}_i+\hat{\bm\Gamma}_i^{\rm T}\bmS_i\hat{\bmGamma}_i)^{-1}
~~\mbox{and}~~
\hat{\bmPsi}=\sum_{i=1}^g\hat{w}_i\mbox{Diag}\{\bmS_i-\hat{\bmB}_i\hat{\bmGamma}_i^{\rm T}\bmS_i\}.
\]
\subsection*{B.7: Model UUC}
For model UUC we have $\bmPsi_i=\psi_i\bmI_p$. The $Q^{[2]}$-function can be expressed as
\[
Q^{[2]}=\sum^{g}_{i=1}\frac{\hat{n}_i}{2}\Big[p\log|\psi_i^{-1}|-
\psi_i^{-1}\mbox{tr}\{\bmS_i\}+2\psi_i^{-1}\mbox{tr}\{\bmB_i\hat{\bm\Gamma}_i^{\rm T}\bmS_i\}-\psi_i^{-1}\mbox{tr}\{\bmB_i(\hat{\bm\Omega}_i+\hat{\bm\Gamma}_i^{\rm T}\bmS_i\hat{\bm\Gamma}_i)\bmB_i^{\rm T}\}\Big],
\]
where $\hat{\bmGamma}_i=(\hat{\bmB_i}\hat{\bmB}_i^{\rm T}+\hat{\psi}_i\bmI_p)^{-1}\hat{\bmB}_i$ and $\hat{\bmOmega}_i=\bmI_q-\hat{\bmGamma}_i^{\rm T}\hat{\bmB}_i$.
Differentiating $Q^{[2]}(\bm\Theta_2|\hat{\bm\Theta})$ with respect to $\bmB_i$ and $\psi_i^{-1}$, respectively, and set the results equal to zero, we obtain
\[
\hat{\bmB}_i=\bmS_i\hat{\bm\Gamma}_i(\hat{\bmOmega}_i+\hat{\bmGamma}_i^{\rm T}\bmS_i\hat{\bmGamma}_i)^{-1}
~~\mbox{and}~~
\hat{\psi}_i=\frac1p\mbox{ tr}\{\bmS_i-\hat{\bmB}_i\hat{\bm\Gamma}_i^{\rm T}\bmS_i\}.
\]
\subsection*{B.8: Model UUU}
For model UUU, there are no constraints, so that
\[
Q^{[2]}=\sum_{i=1}^g\frac {\hat{n}_i}{2}\Big[\log|\bm\Psi_i^{-1}|-\mbox{tr}\{\bm\Psi_i^{-1}\bmS_i\}+2\mbox{tr}\{\bm\Psi_i^{-1}\bmB_i\hat{\bm\Gamma}_i^{\rm T}\bmS_i\}-\mbox{tr}\{\bm\Psi_i^{-1}\bmB_i(\hat{\bm\Omega}_i+\hat{\bm\Gamma}_i\bmS_i\hat{\bm\Gamma}_i^{\rm T})\bmB_i^{\rm T}\}\Big].
\]
Differentiating $Q^{[2]}(\bm\Theta_2|\hat{\bm\Theta})$ with respect to $\bmB_i$ and $\bm\Psi^{-1}_i$, respectively, and set the results equal to zero, we obtain
\[
\hat{\bmB}_i=\bmS_i\hat{\bmGamma}_i(\hat{\bm \Omega}_i+\hat{\bmGamma}_i^{\rm T}\bmS_i\hat{\bmGamma}_i)^{-1}
~~\mbox{and}~~
\hat{\bm\Psi}_i=\mbox{Diag}\{\bmS_i-\hat{\bmB}_i\hat{\bmGamma}_i^{\rm T}\bmS_i\}.
\]

\section*{C: Some computational issues}
\subsection*{C.1: Specification of initial values}
\hspace*{.6cm}
We investigate the issue of getting admissible initial values for
the implementation of the AECM algorithm. Essentially, good
initial values of parameter estimates may speed up the convergence
to the global maximum. A simple procedure of automatically
constructing a set of initial values is outlined below.
\begin{description}
\item[1.]  Perform the $k$-means algorithm initialized with a
random seed. Subtract each observation from its initial cluster
means. Then, do a single tFA fit to these ``centering samples".
The resulting ML estimates
$\hat{\bmtheta}=(\hat{\bmB},\hat{\bmPsi},\hat{\psi},\hat{\nu})$
are taken as initial values of constrained parameters, namely
$\hat{\bmB}^{(0)}=\hat{\bmB}$ for the common factor loading
restriction, $\hat{\bmPsi}^{(0)}=\hat{\bmPsi}$ for the
homoscedastic error covariance restriction, and
$\hat{\psi}^{(0)}=\hat{\psi}$ for the isotropic restriction. The
starting values for the dfs are set as $\hat{\nu}_i^{(0)}=50$, for
$i=1,\ldots,g$, corresponding to an initial assumption of PGMM.

\item[2.]  Initialize the zero-one membership indicator
$\hat{\bmZ}^{(0)}_j=\{\hat{z}^{(0)}_{ij}\}_{i=1}^g$ according to
the $k$-means clustering result. The initial values for the mixing
proportions and component locations are then given by
\[
\hat{w}^{(0)}_i=\frac{\sum_{j=1}^n
\hat{z}^{(0)}_{ij}}{n}~~\mbox{and}~~ \hat{\bmmu}^{(0)}_i=
\frac{\sum_{j=1}^n \hat{z}^{(0)}_{ij}\bmy_j}{\sum_{j=1}^n
\hat{z}^{(0)}_{ij}}.
\]
\item[3.] Divide the data into $g$ groups corresponding to
$\hat{\bmZ}^{(0)}_j~(j=1,\ldots,n)$. Compute the sample
variance-covariance matrix, namely $\bmV_i$, of each group.
Following \cite{mcnicholas08}, the initial values for the
unconstrained $\bmB_i$ are set as
\[
\hat{\bmB}^{(0)}_i=\Big[\sqrt{\hat{\lambda}_{i1}}\hat{\bme}_{i1}~~\cdots~~\sqrt{\hat{\lambda}_{iq}}\hat{\bme}_{iq}\Big],
\]
where $\sqrt{\hat{\lambda}_{ir}}$ is the $r$th largest eigenvalue
of $\bmV_i$ and $\hat{\bme}_{ir}$ is the corresponding
eigenvector. Straightforwardly, the initial values for $\bmPsi_i$
and $\psi_i$ are respectively taken as
\begin{equation*}\
\begin{split}
\hat{\bmPsi}^{(0)}_i&={\rm diag}\{\bmV_i-\hat{\bmB}^{(0)}_i\hat{\bmB}^{(0){\rm T}}_i\},\\
\hat{\psi}^{(0)}_i&=\frac{1}{p}{\rm
tr}\{\bmV_i-\hat{\bmB}^{(0)}_i\hat{\bmB}^{(0){\rm T}}_i\}.
\end{split}
\end{equation*}
\end{description}
\hspace*{.6cm}
In practice, multiple modes typically exist on the complete-data
log-likelihood surface. Thus, the algorithm needs to be
initialized with a variety of starting values. This can be done by
performing $k$-means clustering with various random seeds.

\subsection*{C.2: Stopping rule}
\hspace*{.6cm}
To assess the convergence of the algorithm, various stopping
criteria have been proposed in the literature. The most commonly
used stopping criteria are based on lack-of-progress in the
log-likelihood or parameter estimates and such criteria are not
\textit{bona fide} convergence criteria.\\
\hspace*{.6cm}
We apply the Aitken acceleration scheme
to determine the convergence of each AECM algorithm. The
Aitken's acceleration at iteration $k$ is defined by
\[
   a^{(k)}=\frac{l^{(k+1)}-l^{(k)}}{l^{(k)}-l^{(k-1)}},
\]
where for brevity of notation $l^{(k)}$ means the log-likelihood
value evaluated at $\hat{\bmTheta}^{(k)}$. The asymptotic estimate of the log-likelihood at
iteration $k+1$ is given by
\[
    l^{(k+1)}_{\infty}=l^{(k)}+\frac{1}{1-a^{(k)}}(l^{(k+1)}-l^{(k)}).
\]
The algorithm is claimed to
have reached convergence when
$|l^{(k+1)}_{\infty}-l^{(k)}|<\epsilon$, where $\epsilon$ is the
desired tolerance. Unless otherwise stated, $\epsilon=10^{-5}$
herein.

\subsection*{C.3: Model selection}
\hspace*{.6cm}
The widely used Bayesian information criterion can be used to
choose the best member of the PTMM family and the number of
factors. Herein, the BIC is defined as
\[
       \mbox{BIC}=2\ell_{\rm max}-m\log n,
\]
where $\ell_{\rm max}$ is the maximized log-likelihood and $m$ is
the number of free parameters in the model. Accordingly, models
with large BIC scores are preferred.

\end{document}